\documentclass[10pt,twoside]{cpc-hepnp}

\usepackage{multicol}
\usepackage{graphicx}
\usepackage{booktabs}
\usepackage{amssymb,bm,mathrsfs,bbm,amscd}
\usepackage[tbtags]{amsmath}
\usepackage{lastpage}

\begin{document}

\def\Journal#1#2#3#4{{#1,} {#3,} {{\bf #2}:} {#4}}
\def\NCA{ Nuovo Cimento}
\def\NIM{ Nucl. Instrum. Methods}
\def\NIMA{{ Nucl. Instrum. Methods} A}
\def\NP{{ Nucl. Phys.} }
\def\NPA{{ Nucl. Phys. A}}
\def\PLB{{ Phys. Lett.}  B}
\def\PL{{ Phys. Lett.}}
\def\PPSA{{ Proc. Phys. Soc.} A}
\def\PRP{{ Phys. Rep.}}
\def\PRL{ Phys. Rev. Lett.}
\def\PR{{ Phys. Rev.}}
\def\PRD{{ Phys. Rev.} D}
\def\PRC{{ Phys. Rev.} C}
\def\ZP{{Z. Phys.}}
\def\ZPC{{Z. Phys.} C}
\def\EPJ{{Eur. Phys. J.}}
\def\EPJC{{Eur. Phys. J.} C}
\def\ZPA{{Z. Phys.} A}
\def\MPL{{Mod. Phys. Lett.}}
\def\MPLA{{Mod. Phys. Lett.} A}
\def\CPC{Comput. Phys. Commun.}
\def\JHEP{J. High Energy Phys.}

\def\st{\scriptstyle}
\def\sst{\scriptscriptstyle}
\def\mco{\multicolumn}
\def\epp{\epsilon^{\prime}}
\def\vep{\varepsilon}
\def\ra{\rightarrow}
\def\ppg{\pi^+\pi^-\gamma}
\def\vp{{\bf p}}
\def\ko{K^0}
\def\kb{\bar{K^0}}
\def\al{\alpha}
\def\ab{\bar{\alpha}}
\def\be{\begin{equation}}
\def\ee{\end{equation}}
\def\bea{\begin{eqnarray}}
\def\eea{\end{eqnarray}}
\def\nmsz{\normalsize}
\def\CPbar{\hbox{{\rm CP}\hskip-1.80em{/}}}
\def\ua{\uparrow}
\def\da{\downarrow}
\def\qbar{{\bar q}}
\def\ubar{{\bar u}}
\def\dbar{{\bar d}}
\def\sbar{{\bar s}}
\def\qup{{q^\uparrow}}
\def\qdn{{q^\downarrow}}
\def\uup{{u^\uparrow}}
\def\udn{{u^\downarrow}}
\def\dup{{d^\uparrow}}
\def\ddn{{d^\downarrow}}
\def\qbarup{{\bar q}^\uparrow}
\def\qbardn{{\bar q}^\downarrow}
\def\ubarup{{\bar u}^\uparrow}
\def\ubardn{{\bar u}^\downarrow}
\def\dbarup{{\bar d}^\uparrow}
\def\dbardn{{\bar d}^\downarrow}
\def\sbarup{{\bar s}^\uparrow}
\def\sbardn{{\bar s}^\downarrow}

\def\NCA{ Nuovo Cimento}
\def\NIM{ Nucl. Instrum. Methods}
\def\NIMA{{ Nucl. Instrum. Methods} A}
\def\NP{{ Nucl. Phys.}}
\def\ANP{{Adv. Nucl. Phys.}}
\def\CPC{{Comput. Phys. Commun.}}

\title{Determining the strange and antistrange quark distributions of the nucleon}

\author{%
      F.-G. Cao$^{1}$\thanks{Talk given by F.-G. Cao at QNP 09.}
\quad H. Chen$^{2}$
\quad A. I. signal$^{1}$
}
\maketitle

\address{%
1~(Institute of Fundamental Sciences, Massey University, Private Bag 11 222, Palmerston North, New Zealand)\\
2~(School of Physical Science and Technology, Southwest University, Chongqing 400715, People's Republic of China)\\
}

\begin{abstract}
The difference between the strange and antistrange quark distributions, $\delta s(x)=s(x)-\sbar(x)$, and the combination of light quark sea and strange quark sea, 
$\Delta (x)=\dbar(x)+\ubar(x)-s(x)-\sbar(x)$, are originated from
non-perturbative processes, and can be calculated using non-perturbative models of the nucleon.
We report calculations of $\delta s(x)$ and $\Delta(x)$ using the meson cloud model. 
Combining our calculations of
$\Delta(x)$ with relatively well known light antiquark distributions obtained
from global analysis of available experimental data,
we estimate the total strange sea distributions of the nucleon.
\end{abstract}



\begin{multicols}{2}

\section{Introduction}

The strange and antistrange quark distributions of the nucleon are of great interest.
It has been known for some time that non-perturbative processes involving 
the meson cloud of the nucleon may break the symmetry between the strange and antistrange quark distributions.
This asymmetry affects the extraction of  $\sin^{2} \theta_{W}$ from neutrino DIS processes\cite{NuTeV02}.
A precise understanding on the cross-secrion for $W$ production at the Large Hadron Collider (LHC)
depends on the strange sea distributions at small $x$ region.
However, the strange sea distributions are not well determined compared with those for the light quark sea.
The HERMES Collaboration recently presents their measurement of helicity averaged and helicity dependent
parton distributions of the strange quark sea in the nucleon from charge kaon production in deep-inelastic scattering
on the deuteron\cite{HERMES08}.
The severest constrain on the strange and antistrange distributions before the HERMES measurement
comes from the neutrino(antineutrino)-nucleon deep inelastic
scattering (DIS) in which two muons are produced in the final state, i.e. $\nu ({\bar \nu}) + N \ra \mu^{+(-)}+\mu^{-(+)} + X$. Most data for
such processes are provided by the CCFR\cite{CCFR} and NuTeV\cite{NuTeV} Collaborations. 

There are two dominant mechanisms for the quark sea production in the nucleon: (I) gluons splitting into quak-antiquark pairs,
and (II) contributions from the meson-baryond components in the nucleon. While the sea distributions generated
through mechanism (I) can be assumed to be flavour independent (SU(3) flavour symmetric), i.e. $\dbar=\ubar=\sbar$ and
$d_{sea}=u_{sea}=s_{sea}$ and quark-antiquark symmetric, i.e. $\qbar=q$, the sea distributions generated through mechanism (II) 
violate these symmetries. Mechanism (II) provides a natural explanation for the
observed SU(2) flavour asymmetry among the sea distributions, i.e. $\dbar \neq \ubar$\cite{MCM98},
and predicts a strange-antistrange asymmetry\cite{SignalT87,CaoS99}.

Assuming SU(3) flavour symmetry and quark-antiquark symmetry for the sea distributions generated via mechanism (I), we can construct
a quantity 
\bea
\Delta (x)=\dbar(x)+\ubar(x)-s(x)-\sbar(x),
\label{Delta}
\eea
which has a leading contribution from mechanism (II), and can be calculated using non-perturbative models describing that mechanism.
We present a calculation of $\Delta(x)$ in the meson cloud model (MCM)\cite{Thomas83} by considering 
Fock states involving mesons in the pseudoscalar and vector
octets and baryons in the octet and decuplet.
Combining our calculation for $\Delta(x)$ with results for the light antiquark sea distributions from global PDF fits
we can calculate the total strange distribution $S^+(x)=s(x)+\sbar(x)$
and the strange sea suppression factor  $r(x)=S^+(x)/[\dbar(x)+\ubar(x)]$ 

\section{Formalism}

The wave function for the physical nucleon can be written as
\end{multicols}
\ruleup
\bea
|N\rangle_{\rm physical} &=&  \sqrt{Z} \left( |N\rangle_{\rm bare} 
+ \sum_{BM} \sum_{\lambda \lambda^\prime} 
\int dy \, d^2 {\bf k}_\perp \, \phi^{\lambda \lambda^\prime}_{BM}(y,k_\perp^2) 
|B^\lambda(y, {\bf k}_\perp); M^{\lambda^\prime}(1-y,-{\bf k}_\perp)
\rangle \right).
\label{NMCM}
\eea
\ruledown \vspace{0.5cm}
\begin{multicols}{2}
In Eq.~(\ref{NMCM}) the first term is for a ``bare" nucleon, $Z$ is the wave function renormalization constant,
and $\phi^{\lambda \lambda^\prime}_{BM}(y,k_\perp^2)$ 
is the wave function of the Fock state containing a baryon ($B$)
with longitudinal momentum fraction $y$, transverse momentum ${\bf k}_\perp$,
and helicity $\lambda$, and a meson ($M$) with momentum fraction $1-y$,
transverse momentum $-{\bf k}_\perp$, and helicity $\lambda^\prime$.
The probability of finding a baryon with momentum fraction $y$ (also known as fluctuation function in the literature)
can be calculated from the wave function $\phi^{\lambda \lambda^\prime}_{BM}(y,k_\perp^2)$,
\bea
f_{BM/N} (y)  =  \sum_{\lambda \lambda^\prime}
\int^\infty_0 d k_\perp^2
\phi^{\lambda \lambda^\prime}_{BM}(y, k_\perp^2)
\phi^{*\,\lambda \lambda^\prime}_{BM}(y, k_\perp^2).
\label{fBMN}
\eea
The probability of finding a meson with momentum fraction $y$ is given by
\bea
f_{MB/N}(y)  =  f_{BM/N} (1-y).
\label{fMBN}
\eea
The wave functions and thereby the fluctuation functions can be derived from effective meson-nucleon Lagrangians
employing time-order perturbation theory in the infinite momentum frame\cite{HHoltmannSS}.

The mesons and baryons could contribute to the hard scattering processes such as the deep inelastic scattering,
provided that the lifetime of a virtual baryon-meson Fock state is much
longer than the interaction time in the hard process.
The Fock states we consider include
$\left | N \pi \right >, \left | N \rho \right >, \left | \omega N \right >, \left | \Delta \pi \right >, \left | \Delta \rho \right >$, 
$\left | \Lambda K \right >, \left | \Lambda K^* \right >,  \left | \Sigma K \right >$, and $\left | \Sigma K^* \right >$,
\end{multicols}
\ruleup
\bea
x \delta (x) &=&Z \left[  \left( f_{\Lambda K/N}  + f_{\Lambda K^*/N} \right) \otimes s^\Lambda
			 +\left( f_{\Sigma K/N}  + f_{\Sigma K^*/N} \right) \otimes s^\Sigma 
 -\left( f_{K \Lambda/N}+f_{K \Sigma /N} +f_{K^* \Lambda/N}+f_{K^* \Sigma /N} \right) \otimes \sbar^K  \right], 
\label{deltas} \\
x\Delta (x) &=&Z \left\{ \left( f_{\pi N/N}+f_{\pi\Delta/N}+f_{\rho N/N}+f_{\rho\Delta/N}+f_{\omega N/N} \right) \otimes V_\pi \right.
	-  \left( f_{\Lambda K/N}  + f_{\Lambda K^*/N} \right) \otimes s^\Lambda
			 +\left( f_{\Sigma K/N}  + f_{\Sigma K^*/N} \right) \otimes s^\Sigma  \nonumber \\
	& &~~~~ \left.  -\left( f_{K \Lambda/N}+f_{K \Sigma /N} +f_{K^* \Lambda/N}+f_{K^* \Sigma /N} \right) \otimes \sbar^K  \right\}.
\label{Delta_MCM}
\eea
\ruledown \vspace{0.5cm}
\begin{multicols}{2}
In Eqs.~(\ref{deltas}) and (\ref{Delta_MCM}) $\otimes$ denotes the convolution of two functions, i.e. 
$f\otimes g= \int_x^1 dy f(y) \frac{x}{y} g(\frac{x}{y})$.
The calculation details can be found in Refs. \citep{CaoS99,HHoltmannSS,CaoS01,CaoS03,BisseyCS06,ChenCS09}.

The light quark sea distributions are well decided by the global PDF fits to all available experimental data.  Combining the global fit results
for $\dbar(x)+\ubar(x)$ and our calculation for the $\Delta(x)$ we could have an estimation on the strange sea distributions
\begin{equation}
x\left[ s(x)+\sbar(x)\right]= x\left[ \dbar(x)+\ubar(x)\right]_{Fit}-x\Delta(x).
\label{xS+}
\end{equation}

\section{Results}

In Fig.~1 we show our calculated difference between strange and anti-strange quark
distributions with and without including the contributions from 
Fock states involving $K^{*}$ mesons.
We can see that the contributions from $\Lambda K^{*}$ and $\Sigma K^{*}$ are of 
similar magnitude to those from the lower mass Fock states.

The calculated results for $x\Delta(x)$ together with the HERMES measurement\cite{HERMES08}
and the results from MSTW2008\cite{MSTW2008}, CTEQ6.6\cite{CTEQ6.6}, CTEQ6.5\cite{CTEQ6.5S0} and
CTEQ6L\cite{CTEQ6.5} are shown in Fig.~2.
The HERMES data for $x\Delta(x)$ are obtained by using HERMES measurement for $x S^+(x)$ which is a
leading-oder analysis and CTEQ group's PDFs for $x(\dbar+\ubar)$ at the leading-order, .i.e. CTEQ6L.
The shaded area represents the allowed range for the 
$xS^+$ distribution estimated by the CTEQ group\cite{CTEQ6.5S0} by applying the $90\%$ confidence criteria on the dimuon production data sets,
i.e. by requiring the momentum fraction carried by the strange sea to be in the range of $0.018< \langle x\rangle <0.040$.
It can be seen that our calculations are much smaller that that given in the MSTW2008, CTEQ6L and the central values of the CTEQ6.5
for the region of $x < 0.2$
while the agreement with the HERMES results are reasonably well except for the region around $x \sim 0.10$.
The calculation results agree with that obtained using the CTEQ6.6 PDF set.
It is noticed that our calculations for $x\Delta(x)$ are independent of any  global PDF sets for the proton.
The agreement between our calculations and the CTEQ6.6 results is remarkable. 

\begin{center}
 \includegraphics[width=65mm]{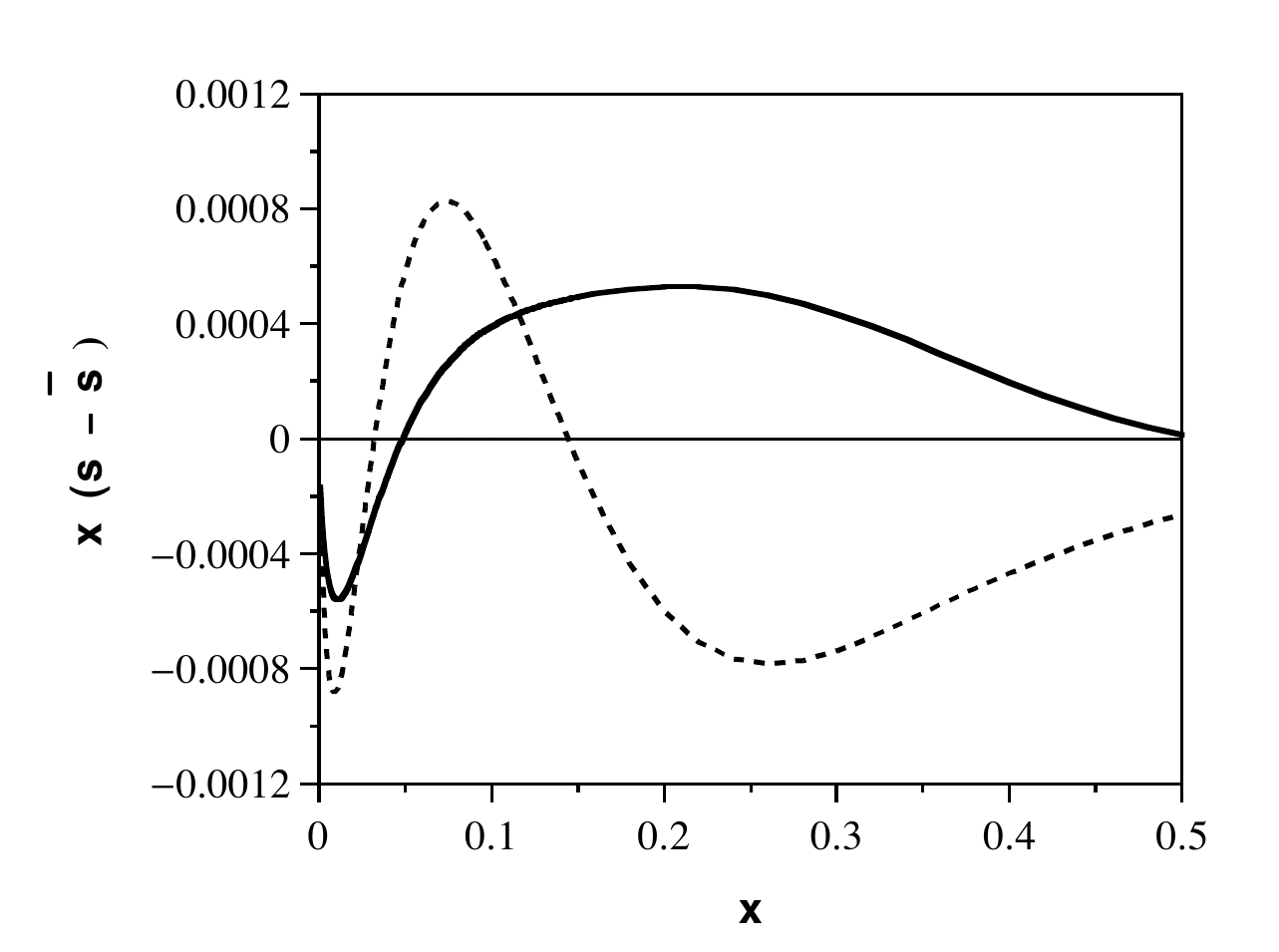}
\figcaption{\label{fig1} The strange sea asymmetry $x[s(x) - \sbar(x)]$  calculated in the meson cloud model.
The solid and dashed curves are the results without and with $K^*$ contributions respectively.  $Q^2 = 16$~GeV$^2$.}
\vspace{-0.0cm}
\end{center} 

\begin{center}
 \includegraphics[width=65mm]{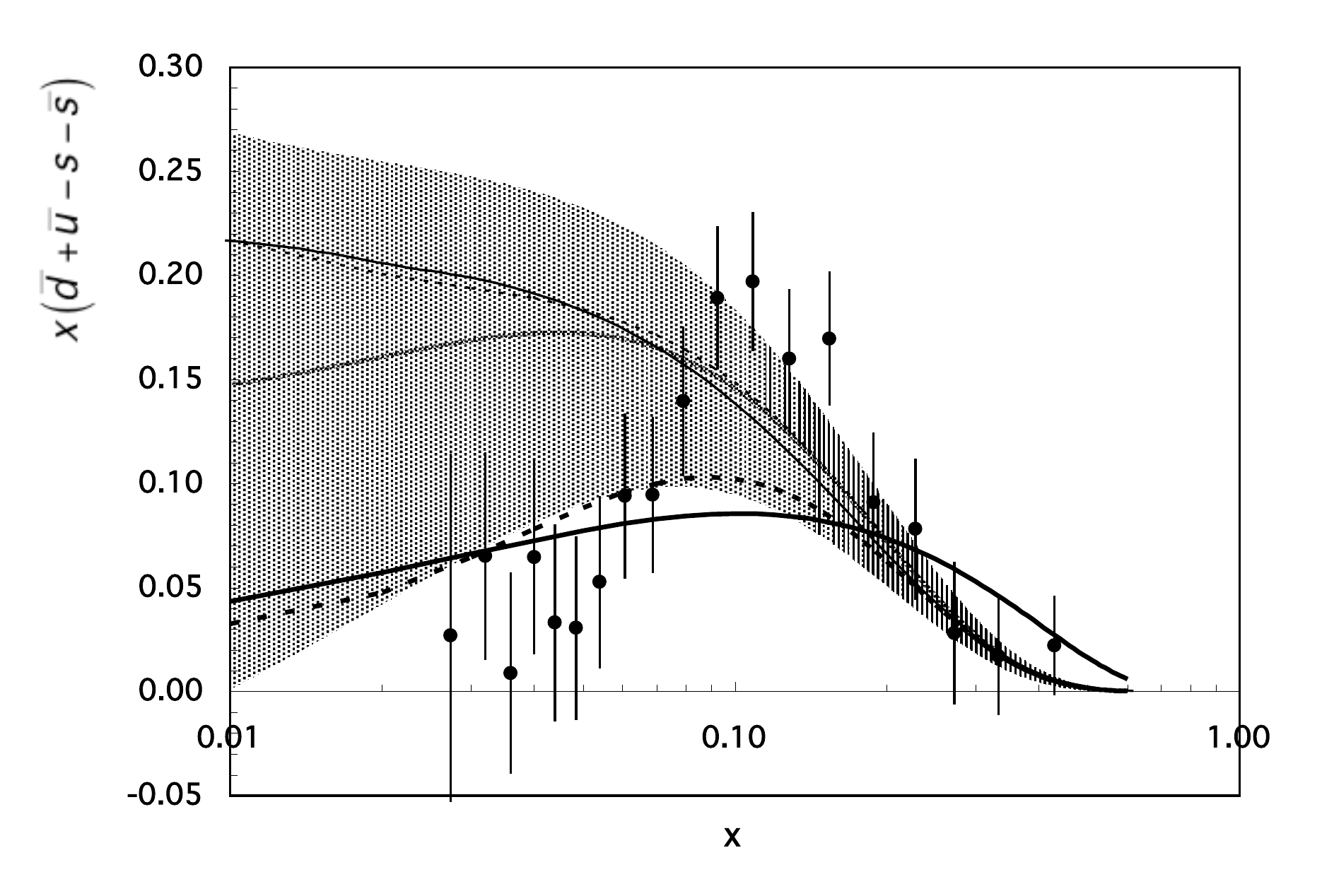}
\figcaption{\label{fig2}A comparison of $x\Delta(x)$. The MCM calculations (the thick solid curve); 
the results obtained using HERMES measurements for $x(s+\sbar)$ and CTEQ6L for $x(\dbar+\ubar)$ (the data points);
the results obtained using NLO analysis of NuTeV data for $x(s+\sbar)$  
and CTEQ6M for $x(\dbar+\ubar)$ (the solid curve); and
the results from MSTW2008 (the dashed curve), CTEQ6.5 (the shaded area), and CTEQ6.6 (the thick dashed curve). $Q^2 = 2.5$~GeV$^2$.}

\vspace{-0.0cm}
\end{center}

\begin{center}
\includegraphics[width=65mm]{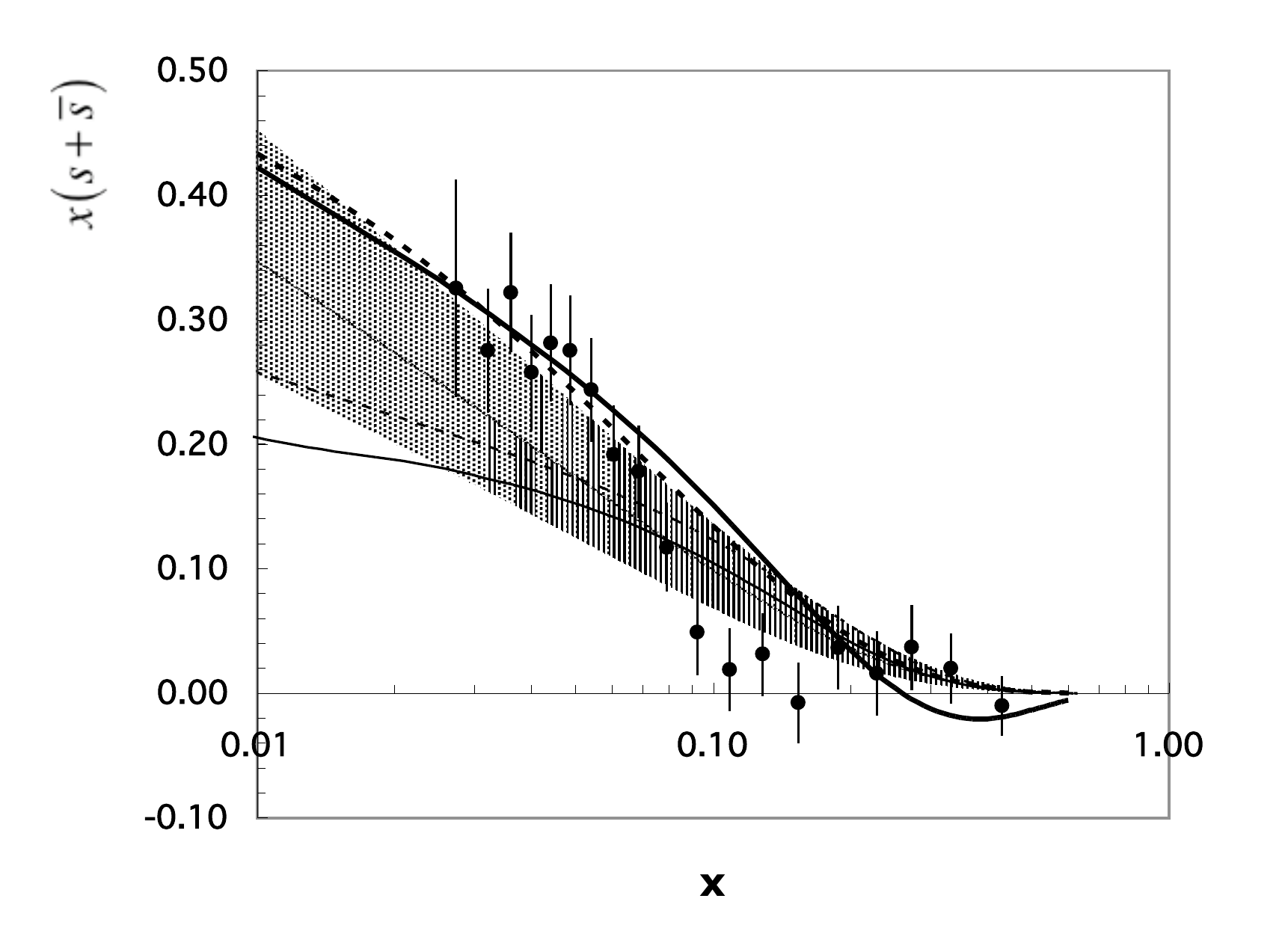}
\figcaption{\label{fig3} The total strange sea distributions, shown as  $x(s+\sbar)$, from the MCM calculations (the thick solid curve),
 the HERMES measurements (the data points),  and  the global fit results from CTEQ6.6M (the thick dashed curve), 
 MSTW2008 (the dash curve) and CTEQ6.5 (the shaded area), and the next-to-leading order analysis of NuTeV dimuon
 data (the solid curve). $Q^2 = 2.5$~GeV$^2$.}
\end{center}

The results for the total strange and antistrange distributions is given in Fig. 3.
In this study the $\dbar +\ubar$ distribution from the CTEQ6.6 set is used. 
It can be found that our calculations agree with the HERMES data and the results from CTEQ6.6 very well for the region of $x<0.07$,
but are larger that that from the MSTW2008 and CTEQ6.5. 
Our calculations for $xS^+$ becomes negative for $x>0.25$ which is unreasonable. 
The reason for this could be that the model calculations over estimate $x\Delta(x)$ or
$x\left(\dbar(x)+\ubar(x) \right )$ is under estimated in the CTEQ6.6 set, or both.

\section{Summary}
We calculated the difference between the strange and antistrange quark distributions and the difference between
the light antiquark distributions
and the strange and anstistrange distributions using the meson cloud model. We estimated the
total strange and antistrange distributions by combining our calculations for the difference
with the light antiquark distributions determined from global parton distribution functions fits.
Our calculations for the strange sea distributions agree with the HERMES measurements and CTEQ6.6 set but larger than 
that given in CTEQ6.5 and MSTW2008.

\end{multicols}
\clearpage
\end{document}